\pdfoutput=1

\documentclass[11pt]{article}

\usepackage[]{emnlp2021}

\usepackage{times}
\usepackage{latexsym}
\usepackage{graphicx}
\usepackage{enumitem}
\setlist{nosep}

\usepackage[T1]{fontenc}

\usepackage[utf8]{inputenc}

\usepackage{microtype}

\title{(Mis)alignment Between Stance Expressed in Social Media Data and Public Opinion Surveys}

\author{Kenneth Joseph \\
  University at Buffalo \\
  Buffalo, NY, USA \\
  \texttt{kjoseph@buffalo.edu} \\ \And
  Sarah Shugars \\
  New York University \\
  New York, NY, USA \\
  \texttt{sarah.shugars@nyu.edu} \AND
  Ryan Gallagher \and Jon Green \and Alexi Quintana Mathé \\
  Northeastern University \\
  Boston, MA, USA \\
  \texttt{gallagher.r,jo.green,quintanamatha.a@northeastern.edu} \\ \AND 
  Zijian An \\ 
  University at Buffalo \\
  Buffalo, NY, USA \\ 
  \texttt{zijianan@buffalo.edu} \\ \And 
  David Lazer \\
  Northeastern University \\
  Boston, MA, USA \\
  \texttt{d.lazer@northeastern.edu} \\ }

\begin{document}
\maketitle
\begin{abstract}
Stance detection, which aims to determine whether an individual is for or against a target concept, promises to uncover public opinion from large streams of social media data. Yet even human annotation of social media content does not always capture ``stance'' as measured by public opinion polls. We demonstrate this by directly comparing an individual's self-reported stance to the stance inferred from their social media data. Leveraging a longitudinal public opinion survey with respondent Twitter handles, we conducted this comparison for 1,129 individuals across four salient targets. We find that recall is high for both ``Pro’’ and ``Anti’’ stance classifications but precision is variable in a number of cases. We identify three factors leading to the disconnect between text and author stance: temporal inconsistencies, differences in constructs, and measurement errors from both survey respondents and annotators. By presenting a framework for assessing the limitations of stance detection  models, this work provides important insight into what stance detection truly measures.   
\end{abstract}

\section{Introduction}
While surveys have a long history of quantifying public opinion \cite{berinsky2017measuring}, deploying them longitudinally requires considerable resources and may produce measurements which lag behind rapidly evolving events. Particularly during ongoing or developing situations, such as the COVID-19 pandemic and U.S. presidential elections, accurately capturing both current and historical opinion may be crucial to crafting appropriate policy interventions or anticipating political outcomes. In contrast to surveys, stance detection algorithms, which identify whether a given text is in favor or against a target concept, can be easily deployed at scale on social media data \cite{aldayelStanceDetectionSocial2021}, positioning them as a potential supplement or replacement for traditional public opinion surveys \cite{kucuk2020stance}. 

However, stance detection methods have rarely been evaluated in comparison to the public opinion data they promise to replace. Instead,  stance detection methods have almost exclusively been evaluated through annotated social media data, often produced by crowdworkers. 
 When disagreements in stance occur across annotators or tasks, existing stance detection work often chooses to either ignore the disagreements \cite{mohammadSemeval2016TaskDetecting2016a,sobhaniDatasetMultiTargetStance2017,samih2021few}, reconcile them with expert annotators \cite{vandenbergNotMyPresident2019}, and/or account for them by  modeling the annotation \cite{joseph2017constance,vandenbergNotMyPresident2019}.

This leaves two critical, largely open questions for stance detection. First, when do stance annotators agree (\textbf{RQ1})? Annotating stance is a subjective process in which annotators use various signals to come to a conclusion \cite{joseph2017constance}. As has been observed in other areas of NLP, embracing this subjectivity and better understanding where disagreements arise can help us improve how we both construct and evaluate our models \cite{balayn2019designing,pavlick2019inherent}.

Second, how well do human annotations of the stance of individual users map to other means of inferring attitudes, and in particular, to survey responses (\textbf{RQ2})? If human coding of text is consistent with survey responses, then algorithms built on those annotations may plausibly replace survey methods. However, if this mapping is inconsistent, researchers may need to consider how stance detection can \emph{supplement} survey methods, rather than replace them. 

Here then, we aim to understand the degree to which human annotations of stance expressed in social media data are consistent across annotators and with public opinion polling. In cases where these measurements disagree, we deconstruct the likely causes of those disagreements. To do so, we leverage data from a longitudinal public opinion survey of Americans conducted in 2020 and early 2021, where respondents voluntarily provided researchers with their Twitter handle. We focus on survey responses to four topics salient in 2020: the U.S. presidential election, lockdowns in response to the pandemic, the use of face masks, and COVID-19 vaccines. We then sample and annotate tweets from survey respondents on these topics. This allows us to directly compare an individual’s stance as perceived by expert annotators and stance as self-reported on the survey. 

With respect to inter-coder alignment (\textbf{RQ1}), we find that annotators vary significantly in their level of agreement based on the target of the stance detection task and how confident they are in their annotations. Disagreements stem from two major sources: stance targets being too complex to reduce to a traditional three point-scale, and the existence of contradictory signals within the information provided for annotation.
The alignment between annotations and the survey data (\textbf{RQ2}) is impacted by the same factors as inter-annotator disagreement. However, these measures are also affected by temporal differences in when the surveys were taken and tweets were sent, and by differences in what individuals are willing to express on surveys as compared to Twitter.

All together, our work suggests that neither stance detection methods nor surveys should be seen as the ``best'' measurement tool, but rather as complementary constructs in the study of public opinion. As we discuss, this points to new directions for both stance detection methodology and the broader practice of connecting survey and social media data. While, due to stringent privacy restrictions on the survey data used, we cannot release survey data, anonymized annotation data that can be used to replicate figures in the present work are included in the supplemental materials.

\section{Background}

\subsection{Stance Detection and Annotation}

While early efforts to use social media to extract public opinion focused on sentiment analysis  \cite{oconnor2010tweets,mitchell2013geography,conrad2019social}, more recent work has shifted towards stance detection, which more directly aligns with the goal of public opinion modeling \cite{sen2020reliability,mohammadStanceSentimentTweets2017}. In stance detection, the task moves from estimating the positive or negative sentiment of a given text to evaluating whether the authoring individual is for, against, or neutral towards some target concept. In principle, this aligns well with traditional public opinion polling in which respondents self-report along a similar stance scale.

However, the growth of stance detection methods has come with many questions about their validity and foundational premises. \citet{sen2020reliability} showed that a variety of existing stance detection tools for Twitter do not generalize well, even when the target is held constant and test data are reasonably similar to training data. 
\citet{joseph2017constance} showed that how one constructs annotation tasks can significantly impact (supervised) model performance and one's assessment of it. Further, as demonstrated by \citet{shen2021sounds} on the closely related task of inferring political ideology, annotator expertise and subjectivity also play an important role in the quality of annotated data.

The present work complements these prior efforts by delving into other questions of annotator disagreement and inference. Whereas prior work has considered disagreement arising from task differences \cite{joseph2017constance}, or properties of the annotators \cite{shen2021sounds}, we control for both of these factors, taking a single task and a relatively homogenous set of expert annotators. Instead, extending recent work studying prediction on multiple targets  \cite{vandenbergNotMyPresident2019,sobhaniDatasetMultiTargetStance2017}, we study how agreement varies depending on the target selected, and how even within a single task design, annotators can come to rely on distinct subsets of information. Second, prior work has largely only {\it assumed} that it is possible for annotators to accurately infer a user's stance. 
The present work tests this assumption by comparing these annotations to self-report data.

\subsection{Linking Surveys to Twitter}

A growing literature exists linking survey data with Twitter data. Much of this work has focused on the difficulties in linking aggregate measures from social media to aggregate measures from polls  \cite[e.g.][]{oconnor2010tweets,diazOnlineSocialMedia2016,beauchampPredictingInterpolatingStateLevel2017}. One general conclusion from this work is that surveys and social media signals at the macro-level are useful, but in different ways \cite{pasekAttentionCampaignEvents2020,buntainComparingSocialMedia2016}. 
The present work complements this perspective from the micro-level, showing that even within individuals, social media and survey data provide distinct and complementary signals. This individual-level linking extends a growing literature where surveys have been linked to Twitter data at the individual level, e.g. in the study of mental health \cite{ernalaMethodologicalGapsPredicting2019}, news exposure \cite{vragaWhoExposedNews2020}, and sharing behaviors \cite{moslehSelfreportedWillingnessShare2020}.

\section{Data and Methods}

In this section, we provide an overview of the survey and Twitter data used, and how we constructed measures of stance for four different stance targets: Donald Trump, COVID-related lockdowns, face masks, and COVID-19 vaccines. We stress that these two datasets are linked, in that 14\% of survey respondents provided a Twitter handle, and so we can connect a subset of our survey respondents to the tweets that they sent. However, for clarity, we describe them separately below.

\subsection{Survey Data}

\subsubsection{Overview}

Our survey data comes from a regularly conducted state-by-state online survey, the COVID States Project,\footnote{\url{https://covidstates.org}} that has been running since April of 2020 regarding issues at the intersection of the COVID-19 pandemic and U.S. politics. Each wave of the survey includes approximately 20,000 respondents, recruited through PureSpectrum, a professional survey company. The company assigns a unique respondent ID that is persistent across all surveys a respondent takes. This allows us to link an individual's responses across multiple surveys.

At the conclusion of the survey, respondents are invited to volunteer their Twitter handle if they have one, as is common in survey-based approaches to Twitter data collection \cite{hughes2021TwitterComparison}. As this is an open-ended response, not every handle provided is authentically associated with the individual who volunteered it. We therefore took a variety of steps to validate the handles we retained for analysis. We first removed handles that 
consist solely of common names (e.g.  @john, @sarah), or that obviously do not belong to individual people (e.g. @Google, @McDonalds). As a proxy for the latter, we removed all accounts that had more than 100,000 followers. We also removed respondents who provided multiple Twitter handles in different survey waves, and Twitter handles associated with multiple respondents.

\subsubsection{Assessing Survey Stance}

In order to make survey responses comparable to the traditional stance detection task, we constructed 3-point (\emph{Pro}, \emph{Anti}, \emph{Neutral}) stance variables for each examined target. Targets and relevant survey questions were determined prior to any annotation of tweets. Full details on the survey questions used are provided in the Appendix.
We only computed survey-inferred stance for respondents who provided a valid Twitter handle and who answered the relevant questions (note that not all questions were asked on all survey waves).

\paragraph{Trump.} We used two survey questions to assess stance towards Donald Trump, one on voting intentions, and one on vote choice asked afterwards.  We then assigned survey stance as follows:
\begin{itemize}
    \item \emph{Anti-Trump:} said they would/did vote for Biden (60\% of respondents).
    \item \emph{Pro-Trump:} said they would/did vote for Trump (32\%).
    \item \emph{Neutral:} said they were unsure of who they did or were going to vote for (8\%).
\end{itemize}

\paragraph{Masks.}  We used two survey questions to determine stance towards masks, one that gauged perceptions of effectiveness, and the other on mask wearing behavior.  We then assigned survey stance as follows:
\begin{itemize}
    \item \emph{Anti-mask:} said masks were \emph{Ineffective} or said they follow mask wearing \emph{Not at all Closely} (15\% of respondents).
    \item \emph{Pro-mask:} said they were \emph{Closely} following mask-wearing guidelines (66\%).
    \item \emph{Neutral:} answered both questions and did not match the conditions for Anti or Pro (18\%).
\end{itemize}

\paragraph{Lockdowns.} We used four survey questions related to social and economic restrictions on 1) leaving the home, 2) business closings, 3) cancelling large events, and 4) closing restaurants to assess stance towards lockdowns. All questions were asked on a four-point Likert scale, ranging from \emph{Strongly Disapprove} (1) to \emph{Strongly Approve} (4).  For each respondent, we averaged the Likert scale values of their answers and used this value to assign survey stance as follows:
\begin{itemize}
    \item \emph{Anti-lockdown:} had an average Likert score of \emph{Strongly Disagree} (i.e. sum of 4-6) (3\% of respondents).
    \item \emph{Pro-lockdown:}  had an average score of \emph{Strongly Agree} (14-16) (60\%).
    \item \emph{Neutral:} had an average score between \emph{Strongly Disagree} and \emph{Strongly Agree} (7-13) (37\%).
\end{itemize}

\paragraph{Vaccines.} In order to account for the roll out of COVID-19 vaccine availability, we used two survey questions to compute stance towards COVID-19 vaccines, one on whether or not they had gotten the vaccine, and if not, what their intentions were. We then assigned survey stance as follows:
\begin{itemize}
    \item \emph{Anti-vaccine:} Stated they were \emph{Extremely Unlikely} to be vaccinated (14\% of respondents).
    \item \emph{Pro-vaccine:} Had already been vaccinated, or were \emph{Extremely Likely} to be vaccinated (43\%).
    \item \emph{Neutral:} Stated they were \emph{Somewhat Likely}, \emph{Neither Likely or Unlikely}, or \emph{Somewhat Unlikely} to be vaccinated (44\%).
\end{itemize}

\subsection{Twitter Data}

\subsubsection{Overview}

After each survey wave, we used the Twitter API to collect the 3,200 most recent tweets for each valid handle. We continued to collect tweets for these accounts on a daily basis, ensuring that once a survey user was added, all of their following tweets would be gathered. In total, there were 15,160 valid Twitter handles, collected over 15 survey waves ranging from April 2020 to February 2021. Of the 15,160 handles provided by the survey respondents, 7,943 (52.4\%) had sent at least one tweet. 

We then identified respondent tweets that were relevant to the four stance targets described above. To do so, we developed keyword lists for each target.\footnote{Lists are provided in the Supplementary Materials.} Keywords for the three COVID-related targets were drawn from a larger list of COVID-related keywords developed in prior work \cite{gallagher2020sustained} and shown to identify COVID-related tweets with over 90\% accuracy \cite{shugars2021pandemics}. We subsetted that list to keywords relevant to the targets we study here. For the remaining target, Trump, we took the keyword list used by \citet{joseph2017constance} for their study on stance towards Clinton and Trump in 2016 and updated the keywords to focus on the 2020 U.S. presidential election. In total, we collected 29,935 tweets about vaccines from 1,338 respondents, 485,906 tweets about Trump from 3,323 respondents, 26,942 tweets about lockdowns from 2,668 respondents, and 28,609 tweets about masks from 1,816 respondents.

We performed a four-step stratified sampling procedure to ensure broad coverage across different survey responses, targets, and types of Twitter users. In the first step, we linked each tweet with its author's survey responses. If a user had responses from more than one survey wave for a given target, we selected the most recent wave to define that user's stance for this sampling procedure. Second, for each target, we constructed three bins of users based on their survey stance. Third, to avoid over-sampling highly active users, we further binned users based on the frequency with which they tweeted. For each target, we split users into three quantiles of activity based on the number of tweets they sent, giving us a set of Low, Moderate, and High activity users. Finally, we sampled up to 40 users from each of the 36 activity level (Low, Moderate, High) by stance (Anti, Pro, Neutral) by target bins, and sampled a single tweet to annotate for each user. By following this procedure, we ensured an equitable representation of each stance and each activity bin for the sample of tweets of each target. Once identified, all tweets across all bins were randomized for annotation.  

\subsubsection{Assessing Tweet Stance}

We determined stance for tweets via an annotation task. For ethical reasons (see Section~\ref{sec:ethics}), five of the paper authors annotated the tweets, rather than crowdworkers. To ensure a realistic annotation setting, four of the five authors were blind to the sampling procedure described below. The fifth performed the sampling but did not know how many users were in each sampled bucket, blinding them from the distribution of survey responses.

Each annotator labeled approximately 570 tweets, and each tweet was labeled by two annotators; disagreements were resolved by a third author. Annotators were asked two questions per tweet. First, they were asked, ``Given the tweet above, do you believe this user is [\emph{Pro}, \emph{Anti}, or \emph{Neutral}]? Wording for answers then was modified to fit the given target (e.g. \emph{Pro-Vaccine}). For Trump, we followed \citet{joseph2017constance} and asked annotators to equate Pro-Biden sentiments with the Anti-Trump label. Second, annotators were asked how \emph{confident} they were: \emph{Very (Confident)}, \emph{Somewhat}, or \emph{Not at all}. Annotators were asked to use the \emph{Very Confident} label only when stance was explicit in the text; otherwise annotators were asked to infer the user's stance, as is typical in stance annotation \cite{mohammadSemeval2016TaskDetecting2016a}. The annotators were asked to use the \emph{Somewhat} or \emph{Not at all} label to define their confidence in that inference.\footnote{Full instructions to annotators are provided in the Appendix, along with an example annotation.}


%
 
\section{Results}

Overall, we analyze 1,372 tweets from 1,129 Twitter users, linked to a total of 1,961 survey responses across all waves.  For both research questions, our analysis relies on a blend of quantitative findings and descriptive analyses. The latter are drawn from qualitative notes taken by authors during annotation, and solidified via discussion amongst authors after the annotation task was completed. For \textbf{RQ1}, we analyzed variation on agreement across all tweets. For \textbf{RQ2}, we study only tweets on which neither annotator was {\it Not at all} confident, because as discussed below, annotations where at least one annotator was not at all confident had unacceptable levels of agreement to indicate we would be analyzing a coherent label for the text.

\subsection{(Dis)agreement Among Annotators}

\begin{figure*}[t]
    \includegraphics[width=\textwidth]{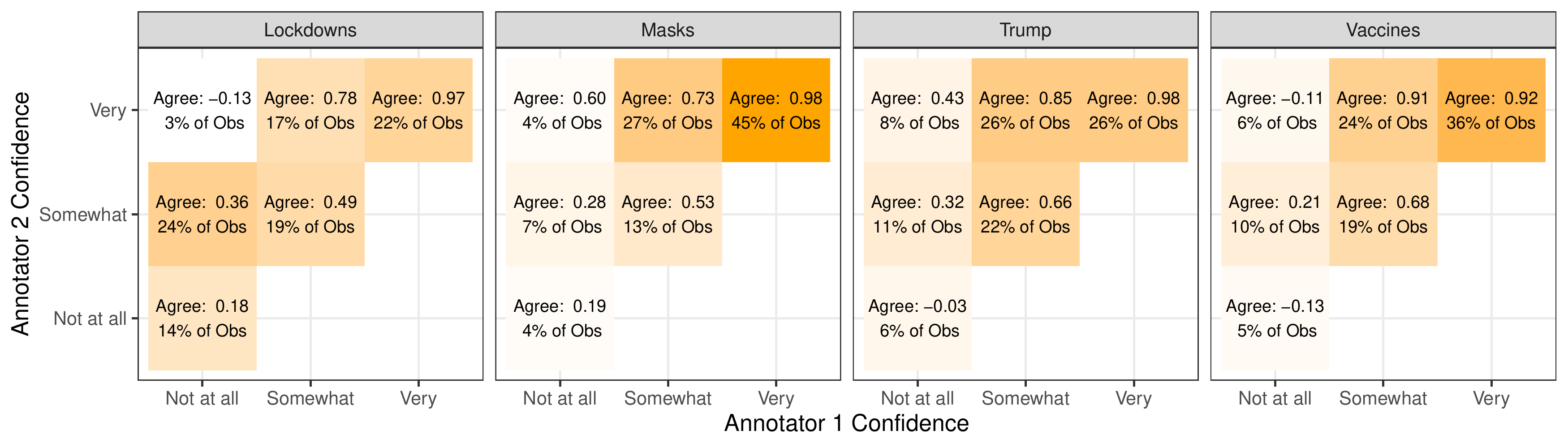}
    \caption{Each cell on each subplot provides agreement statistics (Krippendorf's alpha) and the percentage of annotations for that target accounted for by each combination of annotator level of confidence. Color reflects percentage of all annotations. For plotting ease, Annotator 1's confidence (x-axis) is always less than or equal to Annotator 2's confidence.}
    \label{fig:ann}
\end{figure*}

Annotator agreement varied significantly across targets, and was predominantly impacted by the level of confidence annotators had in their annotations. Figure~\ref{fig:ann} displays agreement statistics (Krippendorf's alpha) for each combination of annotator level of confidence. Annotators agreed on almost every tweet when both were very confident, regardless of target. Even if only one annotator was \emph{Very Confident}, agreement levels were between 0.7-0.9---comparable to prior work that has reported Krippendorf's alpha for stance annotation \cite{joseph2017constance,vandenbergNotMyPresident2019}. Such annotations accounted for well over half of tweets for masks and vaccines (72\% and 60\%, respectively), and around half of Trump tweets (52\%).

However, in the remaining tweets, where neither annotator was {\it Very Confident}, agreement dropped precipitously. For example, across the 61\% of lockdown tweets where neither annotator was \emph{Very Confident}, Krippendorf's alpha was 0.40, well below reliable levels.

Post-annotation discussion surfaced two issues that made annotation difficult. First, as prior work has noted \cite{vandenbergNotMyPresident2019}, off-target tweets, or tweets that were nominally on target but expressed virtually no indications of stance, were difficult to annotate. These occurred for a variety of reasons, the most common being individuals retweeting friends selling homemade masks, and tweets about (often virtual) events happening during lockdown. But all annotators were instructed to use the {\it Not at all} confident label for these cases, and as Figure~\ref{fig:ann} shows, such tweets accounted for a minority of cases in which annotators lacked confidence.

Second, particularly with lockdowns, annotators confronted tweets in which a policy stance was not necessarily reflected in the content or that displayed conflicting stances within the same tweet. For example, 6.4\% of annotated tweets about the lockdowns mentioned homeschooling. The majority of those tweets were parents who expressed exhaustion about homeschooling. While this was a relevant and direct consequence of the lockdowns, many of these same parents also recognized and supported the public health need for lockdowns. For example, one heavily retweeted tweet on Twitter, also retweeted by an individual in our dataset, stated ``Staying at home with kids is more stressful than going to work, according to [a new study].''  While that tweet expresses lockdown fatigue, it does not necessarily convey enough information to determine the stance of an individual in regards to lockdowns as a matter of health policy.

\subsection{(Dis)agreement Between Annotators and the Survey}

\begin{figure}[t]
    \centering
\begin{tabular}{cc}
    \includegraphics[width=.98\linewidth]{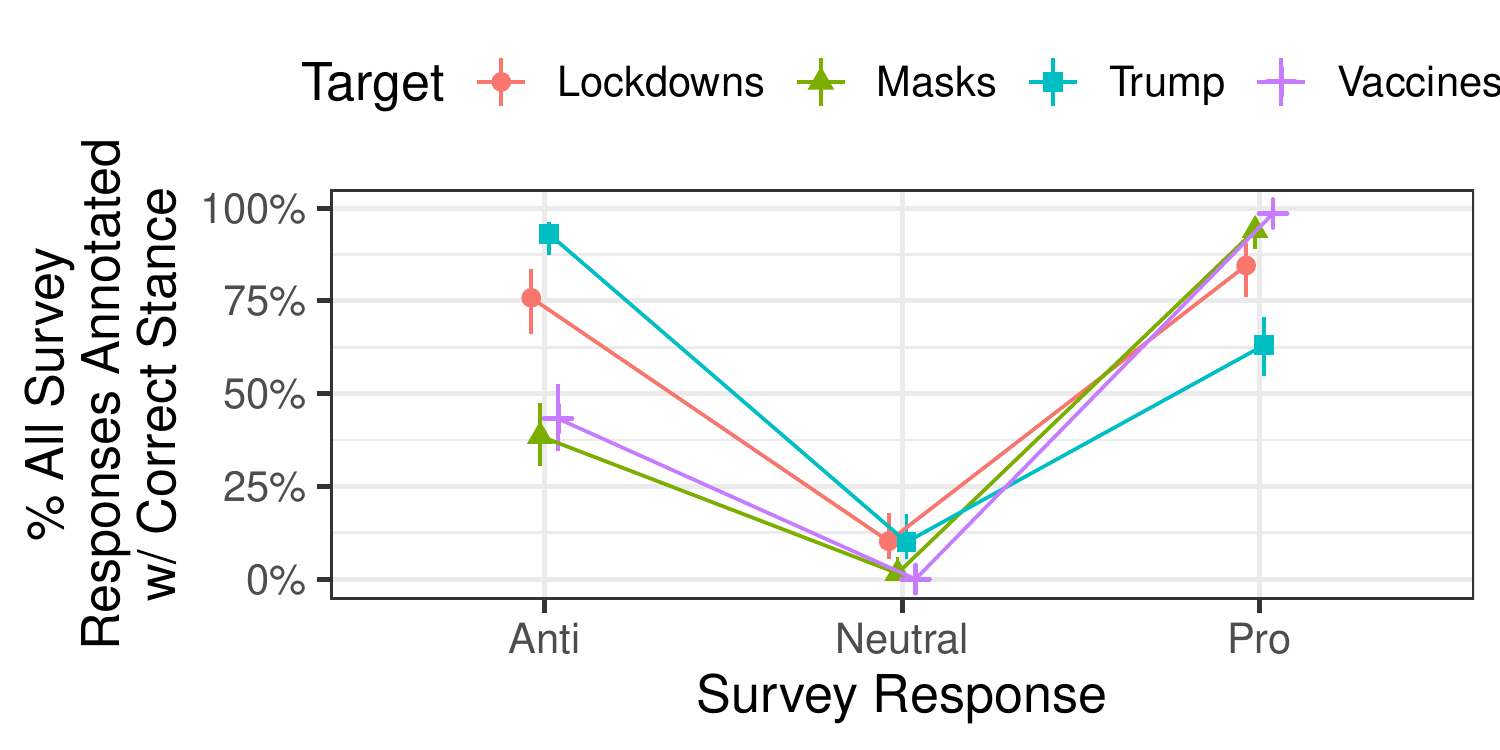} \\
    \includegraphics[width=.98\linewidth]{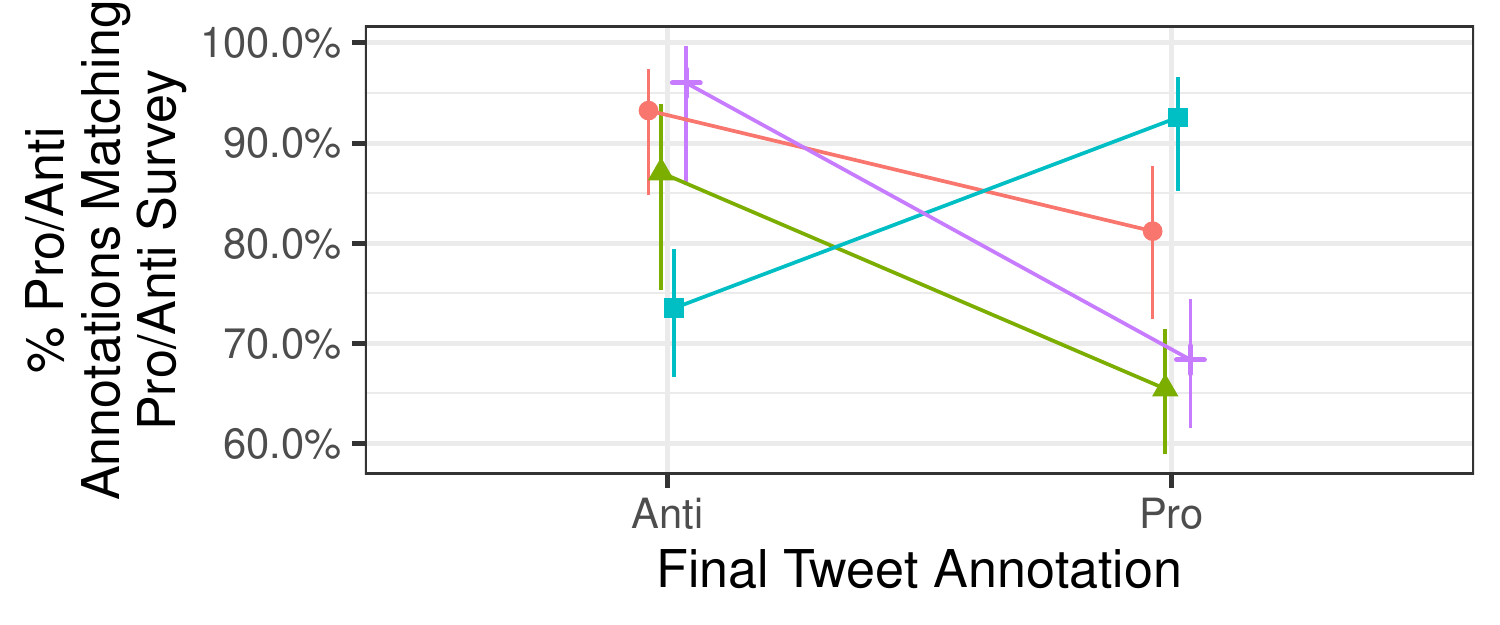}
\end{tabular}
    \caption{Top; Recall of the annotations, treating surveys as ground truth. Bottom; Precision of the annotations, treating surveys as ground truth. Both plots include only tweets where neither annotator was not at all confident. The bottom panel is subsetted to remove \emph{Neutral} responses. }
    \label{fig:intro}
\end{figure}

For some stance/target combinations, annotations and surveys were highly consistent. Specifically, survey respondents who had an \emph{Anti-Trump}, \emph{Anti-lockdown}, \emph{Pro-lockdown}, \emph{Pro-mask}, or \emph{Pro-vaccine} survey stance were almost always annotated with the same stance (top of Figure~\ref{fig:intro}). Hence, recall was high for these stance/target combinations. Ignoring all \emph{Neutral} stances, annotations for \emph{Pro-Trump}, \emph{Anti-masks}, \emph{Anti-vaccines}, and \emph{Anti-lockdowns} matched survey stances over 90\% of the time (bottom of Figure~\ref{fig:intro}). However, given that both the survey and Twitter data correspond to the same individuals, one might \emph{expect} there to be a stronger overlap. Perhaps more interesting are the clear differences that arise between surveys and annotations in the cases not mentioned above. 

The most glaring difference is that annotated data significantly underestimates the number of \emph{Neutral} survey stances. Recall of \emph{Neutral} survey stances was near 0\% across all four targets, because annotators almost never confidently identified \emph{Neutral} stance (top of Figure~\ref{fig:intro}). Thus, the \emph{Neutral} tag was mostly used to express the lack of indicators of stance in the tweet.

\begin{figure}[t]
    \includegraphics[width=\linewidth]{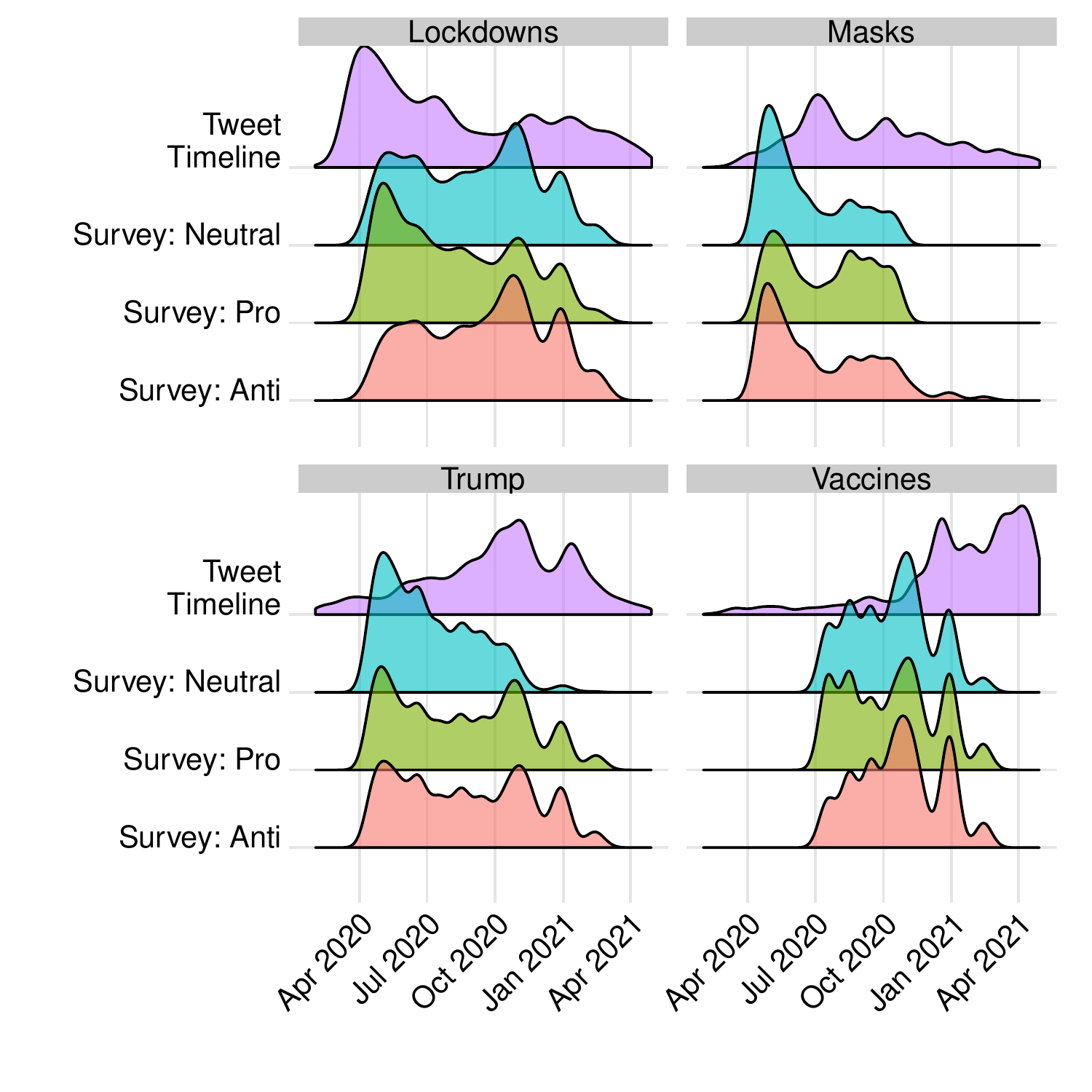} 
    \caption{For each target, density over time of the \emph{Neutral}, \emph{Pro}, and \emph{Anti} survey stances and the number of tweets posted about the target.}
    \label{fig:time}
\end{figure}

One important factor leading to this is that, as shown in Figure~\ref{fig:time}, users tended to tweet about the target topics well after the surveys were taken. In other words, the survey was designed to capture opinions on evolving topics which were emerging in salience. While people were making up their minds at the time of the survey, they may have settled on a stance by the time they felt inclined to tweet about that topic. With respect to masks, for example, the bulk of surveys were collected in May and June of 2020, as guidance and scientific evidence was still evolving.  In contrast, most Twitter activity on the topic began in July, as masks evolved into a partisan flashpoint. With respect to vaccines, surveys were conducted to identify views on vaccines as they were being developed in 2020, but Twitter activity on vaccines picked up only as vaccines began being distributed in November of 2020. Finally, Twitter activity relevant to Trump and Biden naturally peaked around the election and the Capitol Riots, whereas survey data we used was collected throughout 2020. This points to the different roles surveys and stance detection may play in tracking public opinion---surveys can serve as an important indicator of how public opinion will develop while social media can reflect the most vocal opinions at a given moment.

We also see evidence of time as a factor even beyond differences in \emph{Neutral} survey stance. Such differences were large for certain target/stance combinations. Specifically, ignoring all survey stances marked as \emph{Neutral}, around 30\% of users annotated with a \emph{Pro-vaccine} or \emph{Pro-mask} stance provided an \emph{Anti-vaccine} or \emph{Anti-mask} survey response, and around 25\% of users annotated with an \emph{Anti-Trump} based on the social media content provided a {\it Pro-Trump} survey response. Here, again, time seems to have been relevant. For example, most vaccine tweets in our dataset were about individuals receiving their vaccinations; disagreements between annotations and surveys seemed to stem in large part from people who were resistant to vaccines early on but for one reason or another opted to eventually get one.

\begin{figure}[t]
    \includegraphics[width=\linewidth]{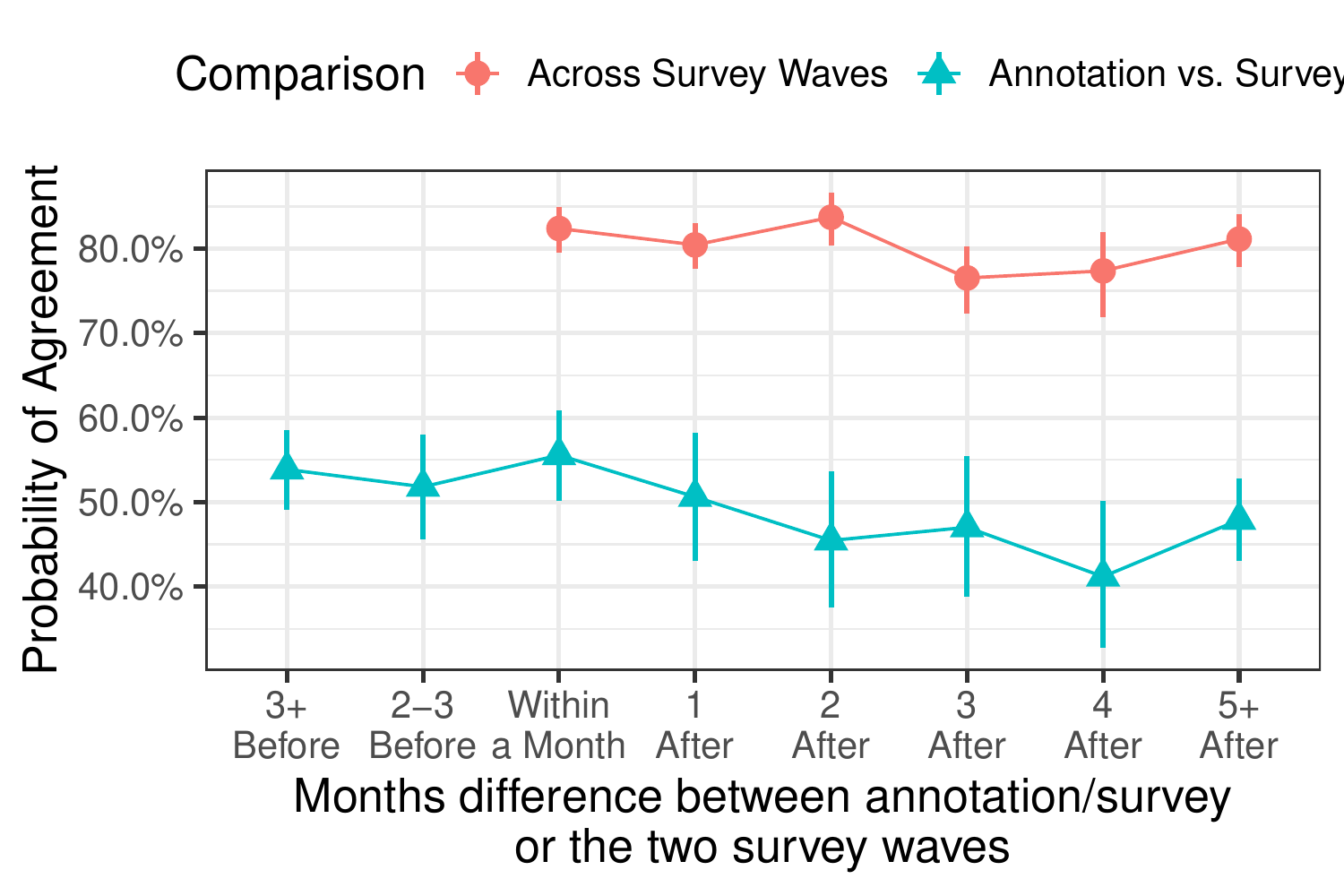} 
    \caption{The probability that for a given survey respondent providing a Twitter handle, their responses on any of the targets across two waves were the same (red circles), or their annotated tweet and survey response(s) were the same (teal triangles). The horizontal axis indicates the time between two survey waves or a survey way and an annotated tweet. 
    }
    \label{fig:time_comp}
\end{figure}

One might be tempted to assume, then, that time was the {\it only} factor differentiating annotated labels from survey responses. If this were the case, however, we would expect the change over time to be similar across stance measures available in our dataset. Specifically, stance change between when the annotated tweet was sent and when the survey was taken should match a similar rate of change in the survey data when individuals took the survey across multiple waves. This would mean the misalignment of stance as annotated in the tweets compared to stance in the survey could be due to the user stance itself changing. Figure~\ref{fig:time_comp} shows this is not the case. While there is evidence that as time increases between survey waves, or between when the annotated tweet was sent and the survey, agreement decreases, the base rate of agreement across survey waves is substantially higher across all time windows. Therefore, while time did have an impact, other, more fundamental differences between surveys and annotations seemed to cause discrepancies between the two.

Again, this may point to fundamental differences in what surveys, versus social media, are measuring. Survey waves took place at regular intervals---every other week---and asked about issues with a range of immediate salience. People tweet, on the other hand, when they have something particularly timely to say. As we see in Figure~\ref{fig:time}, this means that survey responses capture fine-grained opinion at multiple time points, while Twitter stance may be biased towards event-based stance. People are more likely to tweet about targets \emph{during} events like
vaccination drives, elections, and initial implementations of lockdown and mask mandates rather than leading up to them or long after them. Public opinion is, however, still critical during these periods. Furthermore, the bottom of  Figure~\ref{fig:intro} shows that where annotators disagreed with survey data, it was consistently in ways that would align with a liberal viewpoint (i.e. \emph{Anti-Trump}, \emph{Pro-masks}, \emph{Pro-vaccines}, and \emph{Pro-lockdown}). While, time may have played some role here, it is also worth noting that Twitter is more left-leaning overall \cite{shugars2021pandemics}, which may influence which stances people are willing to express on Twitter, relative to the surveys. 

\section{Discussion and Conclusion}

We find important indicators of stance interpretability and how it compares to survey measures. With respect to when annotators agree with each other (\textbf{RQ1}), agreement was very high when at least one annotator was {\it Very Confident}, i.e. when at least one believed stance was explicit. When both annotators were {\it Somewhat} or {\it Not at all Confident}, i.e. when both felt they had to infer stance from indirect statements, agreement between annotators was moderate, low, or even worse than chance. With respect to how well annotations agreed with survey responses (\textbf{RQ2}), the later respondents took the survey after a target was added to the survey, the more likely the respondent was to provide a non-\emph{Neutral} stance. If respondents did provide a non-\emph{Neutral} stance, annotations often matched with the survey data. There were, however, two exceptions to this. First, annotators rarely agreed with surveys for lockdowns, because they struggled to agree with each other. Second, annotators also frequently labeled {\it Anti-Mask} and {\it Anti-vaccine} survey stances as {\it Pro-Mask} or {\it Pro-Vaccine}, respectively, and labeled {\it Pro-Trump} survey responses as {\it Anti-Trump}. We have presented two possible reasons for these latter disagreements between survey and annotation: opinion change and different social norms on surveys as opposed to Twitter. 

Our findings are not without limitation. Our operationalization of survey stance was one of many possible strategies. While the consistency of our main findings across four targets suggests that our overarching points are consistent with the data, choices we made here might nonetheless impact specific findings. Further, we rely on respondents providing accurate handles for themselves, another potential source of error. We performed cleaning to try to ensure this, but our methods likely were not infallible.  

With these limitations in mind, our findings have several implications. First, one oft-stated claim in support of stance detection for social media is that, relative to surveys, it can better capture opinion dynamics.  Our work challenges that assumption at the individual level, because we find that individuals seem to tweet only when their minds are made up, whereas surveys can capture responses as individuals are making up their minds. In any case, many stance detection models leverage features about an individual that are relevant to past behaviors, making implicit assumptions about the static nature of individual stance. For example, a consistent feature in supervised models is who an individual follows \cite{aldayel2019your}, and more recent work has focused on historical retweeting patterns \cite{darwish2020unsupervised}. Even if stance itself were static, we find reason to be wary of assuming the meaning of these features are static as well. An example is an individual who retweeted a pro-mask tweet from Rick Scott, a U.S. senator for Florida, early in the pandemic, which might mean something very different than a retweet of Rick Scott only months later as he shifted to a blend of pro- and anti-mask rhetoric.\footnote{https://floridapolitics.com/archives/345073-rick-scott-skirts-specifics-on-covid-19-jacksonville-mask-mandate/}

Second, and related, is that Twitter may thus counter-intuitively be a bad place to look for opinion change at the individual level. From our analysis, we hypothesize that this is because the social norms on Twitter disincentive individuals from publicly posting the kinds of internal debates they have on surveys that lead them to express more neutral attitudes. Instead, Twitter may be a better place to study public opinion at the ``meso-level'' \cite{berinsky2017measuring}, studying how collective shifts in opinions from people who have already made up their minds change.

Third, our results suggest that there seem to be some targets on which stance detection is bound to be unreliable. Our work suggests the main factor driving this is the complexity of the target in terms of the number of relevant policy aspects. Lockdowns present a useful example for future work. Because they pervaded so much of daily life, stance on lockdowns was complex and nuanced, focusing on everything from homeschooling to Universal Basic Income. The number of survey questions we combined to compute a lockdown stance was itself indicative of this point. Indeed, the way we aligned these survey questions to stance may have impacted our results. However, since prior stance detection work has not compared its output with individual-level opinion polls, there are no established guidelines for doing so. A potential avenue for future work, then, might seek to address aspects of stance in the way we now do for, e.g., sentiment \cite{schouten2015survey} and opinion \cite{wu2015flame} mining.

At an overarching level, our results suggest that stance detection applied to social media data does not capture the same thing that is captured by public opinion polling.  This should not necessarily come as a surprise. Stance detection methods are evaluated by their ability to determine individuals’ attitudes as they are \emph{expressed} on social media and \emph{perceived} by an annotator. In contrast, polling data captures attitudes \emph{solicited} by a researcher  and \emph{self-perceived} by individuals. Our observations here suggest potential merit in leveraging, rather than setting aside, these differences.

\section{Ethics Statement}\label{sec:ethics}

Two important ethical issues arose in the process of completing this work. First, as addressed in detail by \citeauthor{stierIntegratingSurveyData2020}'s (\citeyear{stierIntegratingSurveyData2020}) work, there are important ethical considerations in linking social media data to survey data. The first is anonymity. This arose in determining how to perform the annotation task. Specifically, allocating these annotations to crowdworkers would have required significant anonymization to protect respondent privacy. We were both not confident we could perform full anonymization, and believed that even if we could, we might risk entirely losing the meaning of the text. We therefore decided to perform the annotation task ourselves. It is worth noting, however, that while this took a longer time to do, performing the annotations ourselves ultimately gave us considerably more insight into the data, which we leveraged in our analysis. The second ethical consideration regards replicability. Anonymity extends not only to sharing information with crowdworkers, but with other researchers as well. While we acknowledge that the data we used would be valuable to others, we see it as paramount that we respect the privacy of individuals who volunteered their information to us for research purposes. As such, we only share our own annotations, as well as minimal and heavily anonymized survey data. Further, in this paper, we ensure a balance between anonymity and open science by only quoting paraphrases of specific tweets, rather than the tweets themselves \cite{ayers2018don}. At the same time, we believe we have provided sufficient detail (including code and survey questions) for others to conduct similar studies using the same methods and questions. In this sense, our work cannot be reproduced with the same data (due to the stated privacy concerns), but we believe (and have worked hard to ensure that) it is replicable with the same (exact) methodology.

\bibliography{anthology,custom}
\bibliographystyle{acl_natbib}

\appendix

\section{Additional Details on Survey Stance Construction}

\subsection{Survey Stance for Trump}
The two survey questions asked to identify survey stance towards Trump were one that asked about voting intentions (``If the 2020 U.S. presidential election were held today, which candidate would you vote for?"), and one that asked for who one would voted for after the 2020 elections had occurred ("Which candidate did you vote for in the 2020 U.S. presidential election?").

\subsection{Survey Stance for Masks}
We used two survey questions to identify survey stance towards masks. The first gauged perceptions of effectiveness and asked, ``To the best of your knowledge, are [masks effective or not in] preventing a coronavirus (COVID-19)?'' Responses to this question were on a 3-point scale: \emph{Ineffective}, \emph{Effective}, or \emph{Not Sure}. The second question targeted mask-wearing behavior, asking, ``In the last week, how closely did you personally follow the health recommendations [to wear] a face mask when outside of your home?'' This question was asked on a four-point Likert scale, ranging from \emph{Not at all Closely} to \emph{Very Closely}. 

\subsection{Survey Stance towards Vaccines}
We used two questions to identify survey stance towards vaccines. The first asked whether or not they had already been vaccinated. Respondents who were not yet vaccinated were then asked, ``If a vaccine against COVID-19 was available to you, how likely would you be to get vaccinated?'' Responses to this question were on a 5-point Likert scale.

\subsection{Survey Stance towards Lockdowns}
   We used four survey questions related to social and economic restrictions to assess stance towards lockdowns. All questions were prompted by the header ``Do you approve or disapprove of the following measures which federal, state, and local governments could take to prevent the spread of coronavirus (COVID-19) in the next 30 days?'' and were asked on a four-point Likert scale, ranging from \emph{Strongly Disapprove} (1) to \emph{Strongly Approve} (4).  The specific measures we asked about were ``Asking people to stay home and avoid gathering in groups,'' ``Requiring most businesses other than grocery stores and pharmacies to close,'' ``Cancelling major sports and entertainment events,'' and ``Limiting restaurants to carry-out only.'' For each respondent, we averaged the Likert scale values of their answers and used this to assign stance.

\section{Further Details on Annotation}

\subsection{Annotation Instructions}
Below, we list the instructions given to annotators:
\begin{itemize}
    \item Retweets are not always marked as retweets. E.g. You might see a tweet from the New York Times. When tweets do not appear that they are coming from a person, then, treat them as though they were retweeted by a person (they were)
\item Similarly, some replies seem to not link back to the original tweet. This appears to be when a reply in the chain has been deleted. If something seems like an out-of-context statement, you may want to click on that tweet
\item You should always make a best effort to determine the stance of the individual, even if this means making an educated guess.  Use the confidence scale to mark how certain you are. For the confidence scale, you should use the values as follows
\begin{itemize}
\item Very (confident) - You are certain, there is a direct and/or obvious expression of stance. You should otherwise try to avoid using this marker.
\item Somewhat (confident) - There is no direct indication of stance, but you can be reasonably sure that the inference you’re making is correct. That is, you might be able to draw an inference that someone is “Pro-Vax”, even if they only state a Pro-Mask stance, because you know that these two are likely to imply each other.
\item Not at all (confident) - The tweet says nothing about the user’s stance, e,g. the user or the content actually presented in the tweet. You must infer it but you are very unsure about that inference, e.g., you are doing it based on the user’s picture. Note that Not at all confident still requires that you try to infer the user’s stance
\end{itemize}
\item Equate Anti-Trump with Pro-Biden, unless the critique is coming from an obvious conservative, in which case you may opt to select Undecided. Here, we would expect you to be not “Very” confident. The same goes for Anti-Biden, except from the left
\item Vaccine stance in our study refers specifically to COVID-19 Vaccines
\end{itemize}

\subsection{Sample Annotation}
\begin{figure*}
    \centering
    \includegraphics[width=\textwidth]{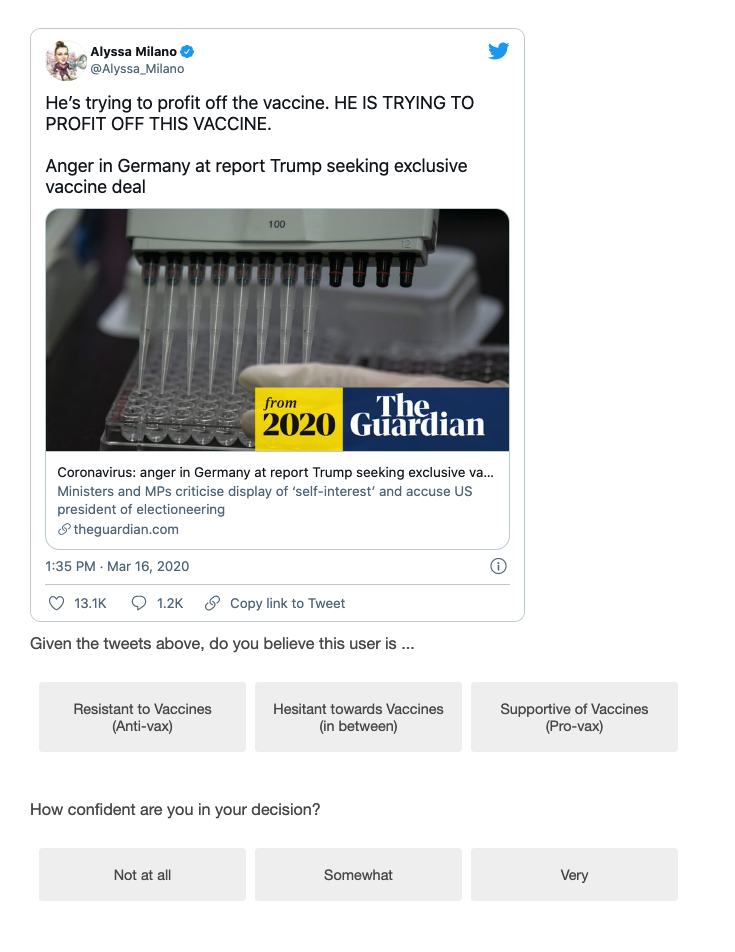}
    \caption{Sample annotation for Vaccines}
    \label{fig:my_label}
\end{figure*}
Figure~\ref{fig:my_label} presents a sample annotation of a very popular tweet that was retweeted by one of the survey respondents (details on the respondent are not shown).
\end{document}